\documentclass[twocolumn,superscriptaddress,showpacs,amsmath,amssymb,pra,longbibliography]{revtex4-1}

\usepackage[pdftex]{graphicx} % Include figure files
\usepackage{dcolumn}  % Align table columns on decimal point
\usepackage{bm}       % bold math
\usepackage[usenames,dvipsnames]{color}
\definecolor{URLCOL}{rgb}{0,0.52,0.83} %external link color
\definecolor{LINKCOL}{rgb}{0.05,0.5,0} %internal link color
\definecolor{orange}{rgb}{0.6,0.3,0} %internal link color
\definecolor{CITECOL}{rgb}{0.25,0,0.48} %link to bibliography
\usepackage{epstopdf}

\usepackage[bookmarks,breaklinks,bookmarksopen,bookmarksnumbered,colorlinks,linkcolor=LINKCOL,linktocpage,citecolor=CITECOL,urlcolor=URLCOL,pdfpagemode=UseOutline,pdftex]{hyperref}

\usepackage[normalem]{ulem}
\usepackage{booktabs}
\usepackage{natbib}

\graphicspath{{f/}}

\definecolor{TITLECOL}{rgb}{0.1,0.2,0.7} %title color
\definecolor{SECOL}{rgb}{0.1,0.2,0.7} %sec color
\definecolor{CONTENTSCOL}{rgb}{0.1,0.2,0.7} %can choose the table of contents title to have same color as sec
\definecolor{SSECOL}{rgb}{0.25,0,0.48} %ssection color
\definecolor{SSSECOL}{rgb}{0.2,0.08,0.53} %subsubsection color  0.2,0.08,0.53
\definecolor{FINCOL}{rgb}{0.01,0.3,0.07} %subsubsection color  0.2,0.08,0.53

\def\coloredtitle#1{\title{\textcolor{TITLECOL}{#1}}} %title color
\def\coloredauthor#1{\author{\textcolor{CITECOL}{#1}}} %author color

\definecolor{URLCOL}{rgb}{0,0.17,0.43} %external link color
\definecolor{LINKCOL}{rgb}{0.05,0.4,0} %internal link color
\definecolor{CITECOL}{rgb}{0.35,0,0.48} %link to bibliography

\def\sss{\scriptscriptstyle\rm}

\def\bea{\begin{eqnarray}}
\def\eea{\end{eqnarray}}
\def\ben{\begin{equation}}
\def\een{\end{equation}}
\def\benu{\begin{enumerate}}
\def\enu{\end{enumerate}}

\def\bei{\begin{itemize}}
\def\eei{\end{itemize}}
\def\beit{\begin{itemize}}
\def\eit{\end{itemize}}
\def\benu{\begin{enumerate}}
\def\enu{\end{enumerate}}

\def\br{{\bf r}}

\def\x{_{\sss X}}
\def\c{_{\sss C}}
\def\s{_{\sss S}}

\def\xc{_{\sss XC}}

\def\H{_{\sss H}}
\def\ee{_{\rm ee}}
\def\ext{_{\rm ext}}

\def\dn{_\downarrow}

\def\intr{\int d^3r\,}

\def\n{n}
\def\t{^{\tau}}
\def\dv{\Delta v}
\def\dn{\Delta n}
\def\sec#1{\section{\textcolor{SECOL}{#1}}}
\def\ssec#1{\subsection{\textcolor{SSECOL}{#1}}}

\def\dv{\Delta v}
\def\dn{\Delta\n}

\begin{document}

\coloredtitle{Exact thermal density functional theory for a model system:
Correlation components and accuracy of the zero-temperature exchange-correlation approximation}
\coloredauthor{J. C. Smith}
\affiliation{Department of Physics and Astronomy, University of California, Irvine, CA 92697}
\coloredauthor{A. Pribram-Jones}
\affiliation{Lawrence Livermore National Laboratory, 7000 East Avenue, L-413, Livermore, California 94550}
\affiliation{Department of Chemistry, University of California, Berkeley, CA 94720}
\coloredauthor{K. Burke}
\affiliation{Department of Physics and Astronomy, University of California, Irvine, CA 92697}
\affiliation{Department of Chemistry, University of California, Irvine, CA 92697}
\date{\today}

\begin{abstract}
Thermal density functional theory  (DFT) calculations often use the 
Mermin-Kohn-Sham (MKS) scheme, but employ ground-state approximations
to the exchange-correlation (XC) free energy.  In the simplest solvable
non-trivial model, an asymmetric
Hubbard dimer, we calculate the exact many-body energies, the exact Mermin-Kohn-Sham functionals
for this system, and extract the exact XC free energy.
For moderate temperatures and weak correlation, we find
this approximation to be excellent.
We extract various exact free energy correlation components and the
exact adiabatic connection formula.
\end{abstract}

\maketitle

%%%%%%%%%%%%%%%%%%%%%%%%%%%%%%%%%%%%%%%%%%%%%%%%%%%%%%%%%%%%%%%%%%%%%%%%%%%%%%%%%%%%%%%%%%%%%%%%%%%%%%%%%%%%%%%%%%%%%%%%

\sec{Introduction}

Recent decades have seen enormous advances in the use of DFT calculations\cite{HK64} of warm
dense matter, a highly energetic phase of matter that
shares properties of solids and plasmas\cite{GDRT14}.  Materials under
the extreme temperatures and pressures necessary to generate WDM can be
found in astronomical bodies, within inertial confinement fusion
capsules, and during explosions and shock physics experiments\cite{DOE09}.
These calculations are used in the description of planetary cores\cite{MD06,LHR09,KDBL15},
for the development of experimental standards\cite{KD09,KDP15}, for prediction of
material properties\cite{HRD08,KRDM08,RMCH10}, and in tandem with
experiments pushing the boundaries of accessible conditions\cite{SEJD14}.
Because of this growing interest in WDM and thermal systems in general, we seek to better understand thermal DFT using exactly solvable models.

In almost all thermal DFT calculations, a crucial approximation is made:
the exchange-correlation (XC)
free energy in principle depends on the temperature\cite{DT81,PD84},
but in practice is approximated by a standard ground-state
approximation.  Most calculations are for extended systems, and usually
use a generalized
gradient approximation, such as PBE\cite{PBE96}.  These Mermin-Kohn-Sham (MKS)\cite{M65,KS65} calculations
predict several key properties, such as the free energy and density for a given distribution of
the nuclei, and any properties that can be extracted from these, such as equations of state of
materials and Hugoniot shock curves\cite{A04}.   If the exact temperature-dependent XC free energy
were known, such properties would be exact\cite{PPGB14}.
In some cases, 
response properties are extracted from the thermal KS orbitals\cite{HRD08}, which involves
a further approximation.
Although no one has shown that the lack of thermal XC
corrections is a fatal flaw in a given calculation, the pervasive use of this uncontrolled 
approximation is an underlying concern\cite{KST12a} that warrants investigation.

The crucial step that made zero-temperature DFT sufficiently accurate for chemical purposes
was the introduction and testing of generalized gradient approximations about
20 years ago\cite{ES99,PHMK05}. By careful comparison with highly accurate benchmarks
produced either by direct solution of the Schr\"{o}dinger equation or from experiments
with well-controlled errors, the general level of accuracy and reliability of such 
approximations was well documented\cite{GSB97,ZKP98,KPB99}.  
With improved binding energies came the
ability to determine molecular geometries for complex systems.
A similar transformation is occurring in materials science today\cite{JOHC13}.

But no such database or highly accurate results exist for thermal systems.  
It is hard to imagine experimental measurements of energies with
the required accuracy under the relevant conditions, but calculations
should be possible. 
Various Monte Carlo methods have been
developed to study WDM in extended systems\cite{FBEF01,SBFH11,DM12,BCDC13,SGVB15,MD15,GSDB16}. 
There have been multiple results from combining Monte Carlo and DFT
for such cases\cite{M09b,VG15,DM15,DM16}.
But none of these could approach the accuracy needed to invert the Kohn-Sham
equations or extract highly accurate correlation energy components.  For such
purposes, finite molecular systems are often the only ones where sufficient
accuracy can be practically achieved.

The prototype case for electronic structure and chemical
binding is the simplest molecule, H$_2$, and its binding energy curve
at zero temperature is simple to calculate, to study the success
of GGA's near equilibrium\cite{PPP97} and their failures as the bond is stretched\cite{B01}.
But even this system is too difficult to calculate when the electrons
are heated:  Only the mean number of electrons is fixed, and {\em all}
possible electron numbers must be included in evaluating the 
grand canonical partition function.

Here we circumvent this difficulty with the simplest representation
of a diatomic molecule.  In a minimal basis set (one function per atom),
the full Hamiltonian is simply a 2-site Hubbard model to which lattice
DFT applies\cite{CFSB15}.  The severe truncation of the Hilbert space makes exact solution
possible in thermal DFT.  By inverting the MKS equations, we perform
 the first exact calculations of correlation
free energies and their individual components for an inhomogeneous system,
an admittedly crude representation of a chemical bond.
By performing self-consistent
calculations with the exact ground-state exchange-correlation
energy functional for this system, we show that the ground-state approximation
works well,
even becoming relatively exact in the high-temperature limit.
We also illustrate several exact conditions on the correlation energy components.
While such a simplified model cannot be used to test the accuracy of standard
approximations applied in the continuum, such as the local density or generalized
gradient approximations, it does provide a first glimpse at the behaviors of 
correlation energy components as a function of temperature, a subject about which
almost nothing is known outside of the uniform electron gas.

This paper is laid out as follows. In section \ref{gsdimer} 
we review the ground-state
of the asymmetric Hubbard model. In section \ref{thermalDFT} we briefly outline thermal DFT.
Next, in section \ref{thermaldimer} we write out the analytic expressions for the MB and MKS system.
Lastly in section \ref{ZTA} we discuss some results using the ground-state XC functional.

\sec{Background}
\ssec{Ground-state Hubbard Dimer}
\label{gsdimer}

Ref. \cite{CFSB15} is an exhaustive review of the asymmetric Hubbard dimer for the ground-state case. In this section we briefly review the Hamiltonian and the most salient points.
The Hamiltonian is typically written as
\ben
\hat{H} = -t\sum_\sigma \left(\hat{c}^\dagger_{1\sigma} \hat{c}_{2\sigma} + h.c. \right)
+ \sum_i\left(U\hat{n}_{i\uparrow} \hat{n}_{i\downarrow} + v_i \hat{n}_i\right)
\label{dimerH}
\een
where $\hat{c}^\dagger_{i\sigma} (\hat{c}_{i\sigma})$ are electron creation (annihilation) operators and
$\hat{n}_{i\sigma} = \hat{c}^\dagger_{i\sigma} \hat{c}_{i\sigma}$ are 
number operators, $t$ is the strength
of electron hopping between sites, $U$ the Coulomb repulsion when two electrons are on the same site, 
and $v_i$ is the external potential on each site.  Without loss of generality, we choose
$v_1 + v_2 = 0$, 
$\dv = v_2 - v_1$, and denote the occupation difference $\dn = n_2 - n_1$.
All terms in Eq. (\ref{dimerH}) have analogs
in an \emph{ab initio} Hamiltonian\cite{CFSB15}. 
The hopping term plays a role logically analogous to the kinetic energy, the Coulomb repulsion is now ultra-short ranged but otherwise the same, and the on-site potential
serves as the one-body potential. Most importantly the asymmetry is necessary to perform our analysis.
Otherwise the occupation difference would vanish and we could not learn about the function(al) behavior.
We choose units where $2\,t =1$ and we vary $U$ and $\dv$.

The key observation is that repulsion and asymmetry directly compete.
When $U$ dominates over $\dv$ the density, $\dn$, tends towards 0, while in the opposite limit $\dn$ tends towards 2. Additionally $U <\dv$ is the weakly-correlated regime while $U> \dv$ is strongly-correlated. The difference between weak and strong correlation is very well characterized in the symmetric case, where 
an expansion in powers of $U$ converges absolutely
up to $U=4\,t$ and diverges beyond that; similarly, an expansion in $1/U$ converges absolutely only
for $U > 4\,t$.  
Here, we restrict our attention to the weakly correlated regime in order to best mimic typical conditions of thermal DFT calculations.

\ssec{Thermal Density Functional Theory}
\label{thermalDFT}

\def\t{^{\rm \tau}}
\def\tl{^{\rm \tau,\lambda}}
\def\HX{_{\sss HX}}
\def\I{_{\sss I}}

In this section we will briefly review the basics of thermal DFT\cite{M65}.  For a more exhaustive treatment see Ref. \cite{PPGB14}.
We begin with an ensemble in thermal equilibrium connected to
a bath at temperature $\tau$. The free energy may be found from:
\ben
A = \min_\n\left( F[\n] + \int d^3r\, n(\br) v(\br)\right)
\label{Adef}
\een
where $v(\br)$ is the one-body potential, $\mu$ is the chemical potential, and
the minimization is over all positive densities with finite kinetic energy.
The Mermin functional is
\ben
F[\n]=\min_{\Gamma\to\n} \mathrm{Tr} \left\{(\hat T + \hat V\ee - \tau \hat S)\Gamma\right\}
\label{Fdef}
\een
where $\hat T$ is the kinetic energy operator, $\hat V\ee$ the electron-electron repulsion
operator, $\hat S$ the entropy operator,
and the minimization is over all statistical density matrices with density $\n(\br)$.
The average particle number is determined uniquely by $\mu$.  Then one can construct the
MKS equations\cite{PPGB14}
\ben
\left\{-\frac{1}{2} \nabla^2 + v\s\t[\n](\br)\right\} \phi\t_i(\br) = \epsilon\t_i\,\phi_i\t(\br),
\label{MKSeqn}
\een
where
\ben
v\s\t[\n](\br) = v(\br)+v\H[\n](\br) + v\xc\t[n](\br),
\label{vsdef}
\een
and $v\H[\n](\br)$ is just the usual Hartree potential\cite{PYTB14} and
\ben
v\xc\t[n](\br) = \frac{\delta A\xc\t[n](\br)}{\delta n(\br)}.
\label{xcpot}
\een
The density is the sum over all orbitals,
\ben
\n\t(\br) = \sum_i f_i\t\, |\phi\t_i(\br)|^2,
\label{ndef}
\een
where $f_i\t = (1+e^{(\epsilon_i^\tau -\mu)/\tau})^{-1}$ are their Fermi occupations.
Finally, once self-consistency has been achieved, the free energy of the interacting
system is reconstructed as:
\ben
A\t = A\s -U\H[\n] + A\xc^\tau[\n] - \intr n(\bm{r}) v\xc^\tau[\n](\bm{r}).
\een
where $A\s$ is the Kohn-Sham free energy.

If the exact XC free energy density
functional (confusingly, often referred to as simply the XC energy)
were known and used in the MKS equations, then their solution produces the exact
density and free energy (and any other quantity that can be directly extracted
from them).  
However, there are very few cases where we have access to the exact
$v\xc(\br)$.   All practical MKS calculations use some approximation, and
most use a simple ground-state approximation.  To distinguish different
levels of approximation, we write
\ben
A\xc\t[\n] = E\xc[\n] + \Delta A\xc\t[\n],
\een
where $E\xc[\n]$ is the {\em exact} ground-state XC energy, and
$\Delta A\xc\t[\n]$ is the difference in XC free energy from its ground-state
value.  We call this the thermal contribution to $A\xc\t$.
Then, the {\em zero-temperature} approximation (ZTA) is where we ignore
the thermal contribution to $A\xc\t$, i.e.,
\ben
A\xc^{{\rm ZTA},\tau}[\n] = E\xc[\n],
\label{eqZTA}
\een
i.e., we neglect thermal effects, but use the {\em exact} ground-state XC functional.
This allows us to separate thermal from non-thermal XC effects in a completely
well-defined manner.   Of course, in practice, it is only in simple model systems
that one has access to the exact ground-state XC functional.

In this language, most modern QMD calculations can be thought to have made two
distinct approximations.  The first is to make ZTA and ignore thermal contributions.
The second is to use some common approximation for $E\xc[\n]$ within ZTA.  On the
other hand, calculations that use, e.g., thermal LDA, go beyond ZTA, but approximate
both the ground-state and thermal contributions to $A\xc\t[\n]$.

\sec{Analytic results}
\label{thermaldimer}

We apply this technology to the asymmetric Hubbard dimer.  The DFT version of a lattice
model is called site-occupation functional theory (SOFT)\cite{SGN95}
and has the distinct advantage of a truncated Hilbert space.
We can compute every energy for every particle number
and construct exact thermodynamic and DFT components. 
The truncation makes the calculation feasible.  We expect that, for very high temperatures,
the results will not be representative of realistic systems with infinite Hilbert
spaces.

\ssec{Exact many-body solution}

To begin, we calculate the finite-temperature many-body energy and density for the Hubbard dimer.  
Begin with the grand canonical partition function
\ben
Z_{gc} = \sum_{i,N} e^{(\mu N - E_i(N))/\tau}
\label{zgc}
\een
where $E_i(N)$ is the $i$-th energy level of the Hamiltonian
with $N$ particles. The energies for 0 through 4 particles are calculated
explicitly,  yielding the exact partition function.
From that we construct the grand potential, its derivatives, and the free energy in the usual fashion\cite{P11}:
\bea
\Omega &=& - \tau \log(Z_{gc}),~~~~~
S = \left.\frac{\partial \Omega}{\partial \tau}\right|_{\mu}, \label{statmechA}\\
N &=& \left.\frac{\partial \Omega}{\partial\mu}\right|_\tau,~~~~~~~~~~~~
A = \mu N - \Omega.
\label{statmechB}
\eea
We choose half-filling, $\langle N \rangle = 2$, which means $\mu=U/2$ (and $\mu=0$ for the MKS system)\cite{S72b}.
With the partition function 
and Boltzmann factors we can calculate ensemble averages:,
\ben
X = Z_{gc}^{-1} \sum_{i,N} \langle \hat{X}\rangle_{i,N} e^{(\mu N - E_i(N))/\tau},
\label{ensavg}
\een
where $\langle \hat{X}\rangle_{i,N}$ is
the expectation value of a general operator $\hat{X}$ of the $i$-th state for $N$ particles.
Using Eq. (\ref{ensavg}) we compute the exact energy components for the dimer. 
To do this, we calculate the expectation values for each particle number of the quantities of interest
such as $T$, $V\ee$, and $\dn$.
We list in the appendix all the expectation values for the total energies, energy components,
coefficients of the eigenstates, and densities for all the particle numbers.

\begin{figure}[htb]
\includegraphics[width=\columnwidth]{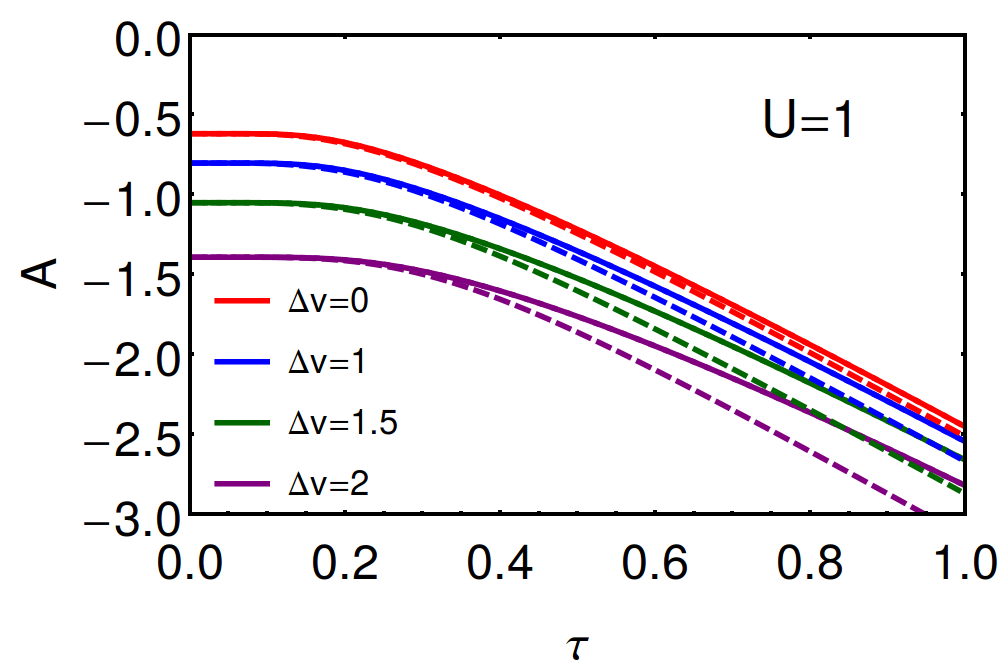}
\caption{Free energy for different values of $\dv$. Solid lines are exact, dashed lines are 
the zero-temperature XC approximation (ZTA), evaluated on the self-consistent thermal density.}
\label{AU1}
\end{figure}

\begin{figure}[htb]
\includegraphics[width=\columnwidth]{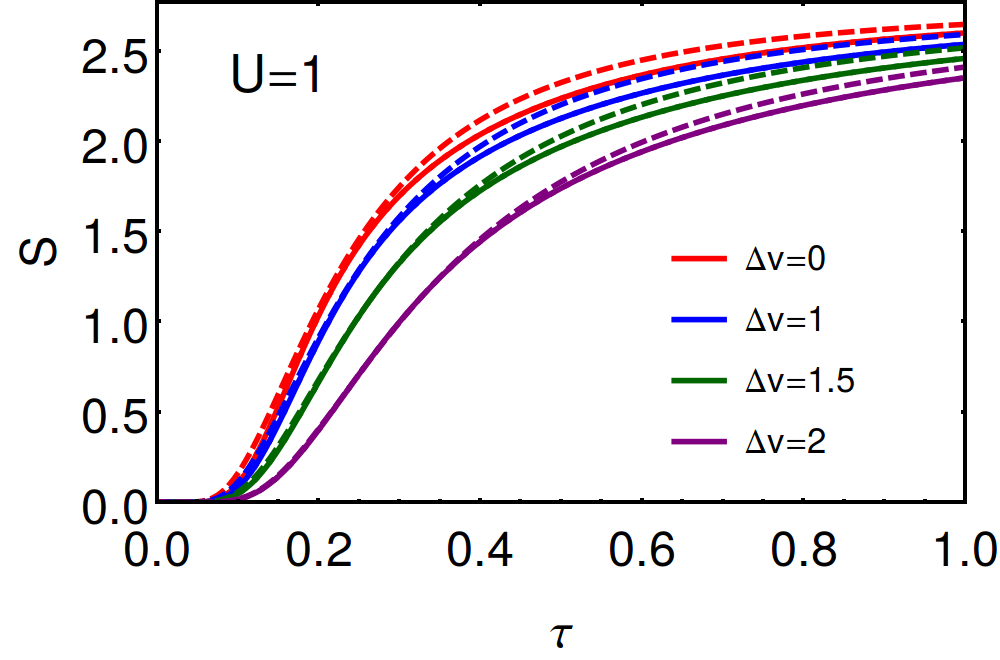}
\caption{Exact entropy (solid) and self-consistent Kohn-Sham entropy (dashed) for different values of $\dv$. All curves approach $4\log 2$.
}
\label{SU1}
\end{figure}

In Fig. \ref{AU1} and \ref{SU1}, we plot the free
 energy and entropy as a function of temperature 
for several different values of $\dv$. For the free energy we include curves for the zero-temperature
approximation and for the entropy we include the self-consistent Kohn-Sham entropy
(both to be discussed later). In both cases we pick a system, i.e. fix $\dv$ and $U$ and see what happens
as we heat it up. For the free energy, the values at $\tau=0$ recover the ground-state energies reported in Ref. \cite{CFSB15}. 
Increasing temperature results in a decrease in free energy primarily due to the entropic term, $-\tau S$, as  expected. 
At small temperatures there is minimal effect as seen in Fig. \ref{SU1} where
the entropy is small and further multiplied by a $\tau \ll 1$ when
calculating $A$. However, once the system
is sufficiently warm the entropy plays a much larger role. 
In contrast, increasing $\dv$ lowers the entropy since the asymmetry restricts the motion of electrons.
Lastly, the entropy approaches
a maximum value of $\log(16)$ for higher temperatures where 16 is the number of states in
our grand canonical ensemble.

\ssec{Inversion and correlation components}

\begin{figure}[htb]
\includegraphics[width=\columnwidth]{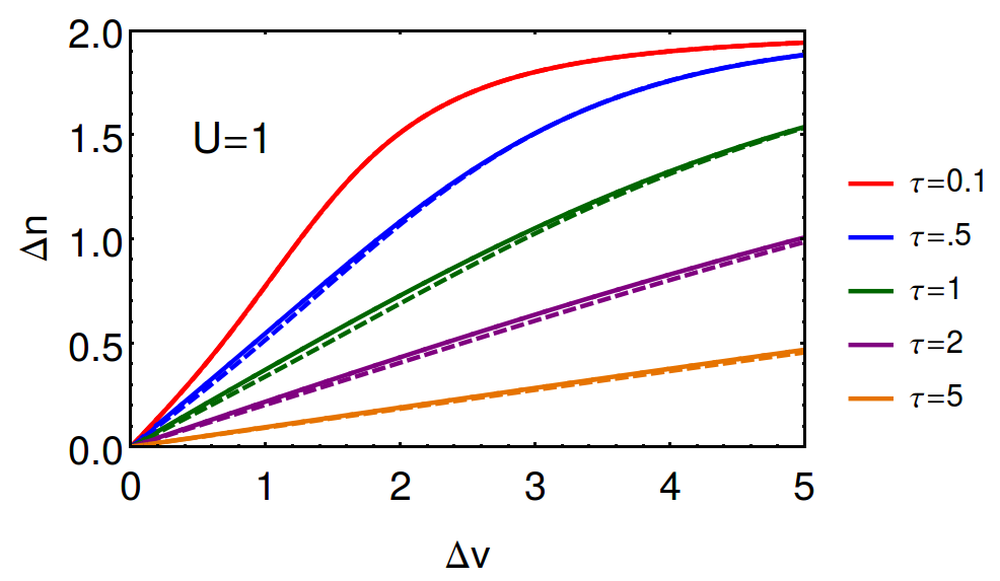}
\caption{Densities as a function of temperature for the
system of Fig. \ref{AU1}. Solid lines are exact, dashed lines
are self-consistent KS using the ZTA.}
\label{dens}
\end{figure}

Next, we construct the exact KS potential
as well as various energy components using the MKS approach.
To begin we construct the exact occupation difference $\dn$ from Eq. (\ref{ensavg}).
We plot
the result in Fig. \ref{dens} for fixed $U$ but against $\dv$ and
vary $\tau$. In this figure we also plot the ZTA result which will be discussed later.
Increasing the temperature pushes the electrons apart, akin to repulsion. 
As the system heats up,
$\dn$ becomes closer to 0 as both electrons sit on separate sites even when $\dv$ is large.

To construct the exact MKS potential, we first give formulas for
non-interacting electrons ($U=0$, a.k.a. tight-binding).

The grand canonical partition function collapses to  the product
\ben
Z_{gc} = \prod_i \left(1 + e^{\beta(\mu-\epsilon_i)} \right)
\een
where $\epsilon_i$ is the single-particle orbital energy. 
Eq. (\ref{statmechA}) and (\ref{statmechB}) can then be used.
The entropy can also be explicitly given in terms of Fermi factors,
\ben
S\s = - \sum_i f_i\log(f_i) + (1-f_i)\log(1-f_i).
\een
where $f_i = (1+ e^{\beta(\epsilon_i-\mu)})^{-1}$.
The Kohn-Sham entropy is calculated in Fig. \ref{SU1} where the Fermi factors
are calculated from self-consistently solving the MKS equations (see below). 

To construct the MKS system for the Hubbard dimer within SOFT, we simply
repeat the exact calculation with $U=0$, i.e., a tight-binding dimer.  
We find:
\ben
\dn =-2 \sin\phi \tanh\alpha
\label{ksden}
\een
where
$\alpha =(4\tau \cos\phi)^{-1}$,
$\sin\phi =x/{\sqrt{1+x^2}}$, $\tau$ is in units of $2\,t$, and
$x=\dv\s/2\,t$.   
To perform the inversion for a given density from the many-body problem,
we perform a binary search at the given temperature on
Eq. (\ref{ksden}) to find $\dv\s(\dn)$, the
exact KS site-potential difference that yields the required occupation density. 
The exact $\dv\xc$
for the given $\dn$ is then found
by subtracting off the other potential contributions, i.e., $\dv$ and $\dv\H$.
The Hartree energy (in the standard DFT definition\cite{PYTB14}) for this model is
\ben
U\H(\dn) = U \left(1+ \frac{\dn^2}{4} \right),
\een
and the Hartree potential is simply
\ben
v\H(\dn) = U \dn/2 
\een
and both functionals are temperature-independent.  For two unpolarized electrons,
$E\x=-U\H/2$ at all temperatures\cite{PPGB14}, and so is also independent of $\tau$.
The thermal MKS hopping energy is just that of this tight-binding problem:
\ben
T\s\t(\dn)/(2t) = \dn/x(\dn)
\een
and the tight-binding MKS entropy is 
\ben
S\s\t(\dn)=4 \log\left\{2 \cosh\alpha \right\} - 4 \alpha \tanh\alpha
\een
With these simple results, we can now extract the correlation free energy for this problem as
\ben
A\c\t = (T\t-T\s\t) - \tau (S\t- S\s\t) + (V\ee\t - U\HX)
\label{Acdef}
\een
where $T\t$, $S\t$, and $V\ee\t$ are calculated from the many-body problem via
eqs. (\ref{ensavg}), (\ref{statmechA}), and (\ref{ensavg}).
Since $A\x\t$ is trivial and has no thermal contribution for our system,
$A\c\t$ is what we study, and
we know of no other exact calculation of this quantity for a finite system.

\sec{Numerical results}

Performing the inversion to explicitly
analyze the MKS potential shows how the features of interactions
are built into the non-interacting potential\cite{TGK08,CCD09,NRL13}.
The crux of the MKS approach is that we capture the effects of interactions
through the modified external potential $\dv\s$. For example,
interaction causes the dimer occupations
to be more symmetric, thus $\dv\s < \dv$ for a MB system with $U>0$.
Similarly, for any given density both potentials, $\dv$ and $\dv\s$, increase
with temperature to counteract thermal effects pushing
the system towards symmetry. 
But even in this simple model, there is a vast parameter space to be explored
as, choosing $2\,t=1$, we can vary $U$, $\dv$, $\tau$, and $\langle N \rangle$.
We focus on $\langle N \rangle =2$, and the weakly-correlated and
low temperature corner of our parameter space:  $U,\tau < 1$.
In particular, we avoid warming our model so much
that properties are strongly influenced by the very limited Hilbert space.
Specifically, we check that the system is
not too hot by computing the occupations of all the 
states in the grand canonical ensemble. We test this in the symmetric case because it is most prone to overheating since asymmetry competes against thermal effects. For $U=1$, uniform
occupation of all states does not occur until $\tau \gg 8$ and appreciable uniformity does not start to arise until $\tau\approx4$. Thus our results are not limited by the top of our Hilbert space.

\begin{figure}[htb]
\includegraphics[width=\columnwidth]{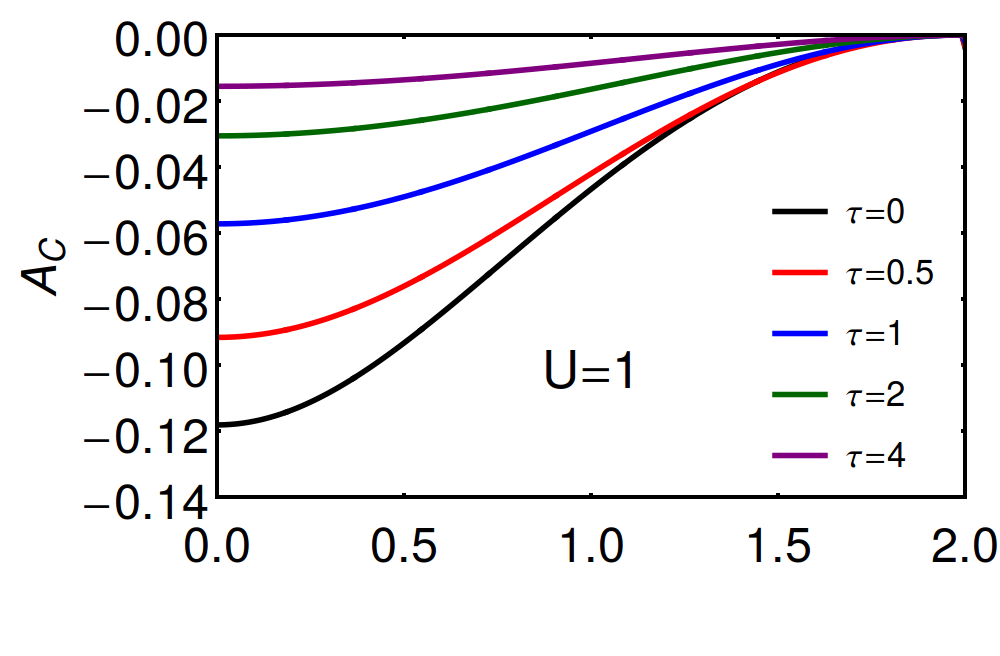}

\vspace{-2em}

\includegraphics[width=\columnwidth]{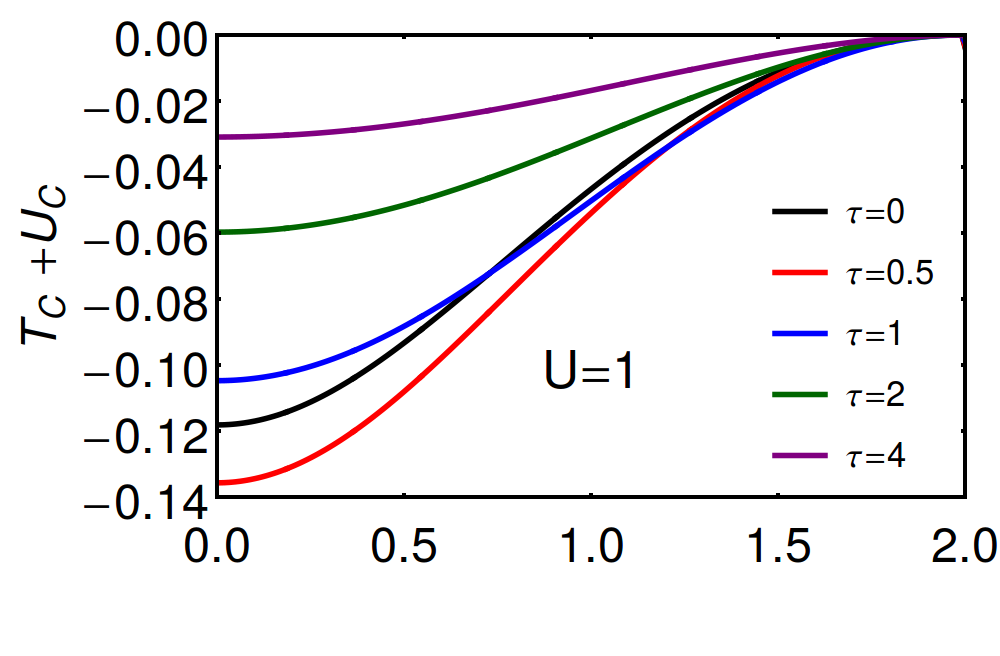}

\vspace{-2em}

\includegraphics[width=\columnwidth]{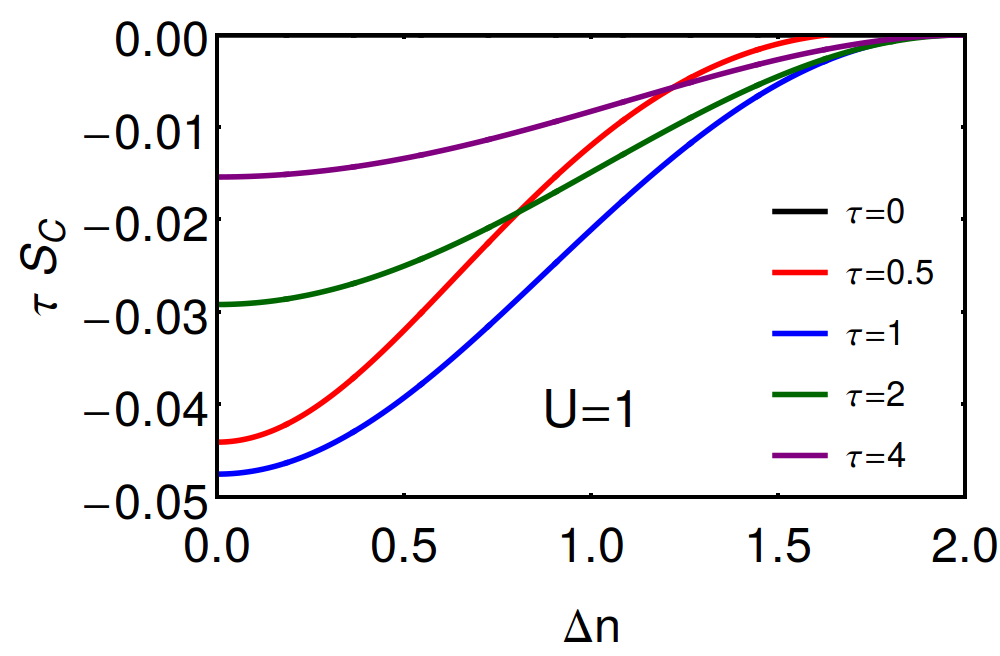}
\caption{
Panel 1: Correlation free energy functional for various temperatures.
Panel 2: Sum of kinetic and potential energy functional for various temperatures.
Panel 3: Entropic correlation functional for various temperatures. 
}
\label{corr}
\end{figure}

We can calculate all the individual contributions to the correlation free energy
by subtracting MKS quantities from their physical counterparts.
These are the energy differences appearing in Eq. (\ref{Acdef}):
\ben
T\c\t=T\t-T\s\t,~~~S\c\t=S\t-S\s\t,~~~~U\c\t=V\ee\t-U\HX.
\label{TSUcdef}
\een
The kentropic correlation is $K\c=T\c-\tau\, S\c$ and plays a key role in
thermal DFT\cite{PPFS11}.
In Fig. \ref{corr}, we plot the exact correlation free energy functional, the sum of kinetic and potential correlation functional, and lastly the entropic correlation functional all for various temperatures. 
By fixing $U$ and $\tau$ and plotting versus $\dn$, we analyze
the correlation as a density functional, i.e. we are no longer looking at a fixed system and
instead are looking at the underlying structure of how thermal DFT behaves.

We see that the correlation free energy is always negative, 
the kentropic contribution is always positive (not shown), and the potential contribution is always negative.
These are consistent with  conditions on the correlation\cite{PPFS11}. This is
the first exact investigation of those inequalities.
The correlation free energy, $A\c\t$, always decreases with temperature 
at $U=1$, even though the
components do not behave that way at small temperature. 
$T\c + U\c$ and $\tau S\c$ also decrease for all densities at
larger temperature just like $A\c$.
In this regime, thermal effects dominate over interactions, resulting in
the interacting system and the non-interacting system having similar
energy components and thus relatively smaller correlation. 
But for small temperature, i.e. $\tau < 1$ when $U=1$, the MKS quantities are furthest
from the exact system since neither effect dominates and this
results in an even larger difference between
the two systems than at $\tau=0$.
Overall we see the same behavior as 
in the ground-state case\cite{CFSB15} -- correlation decreases as our system becomes more asymmetric. If
the electrons are completely pinned on the lower site then there is no motion, the interaction
is completely described by the Hartree, and there is only one entropic conformation.

\begin{figure}[htb]
\includegraphics[width=\columnwidth]{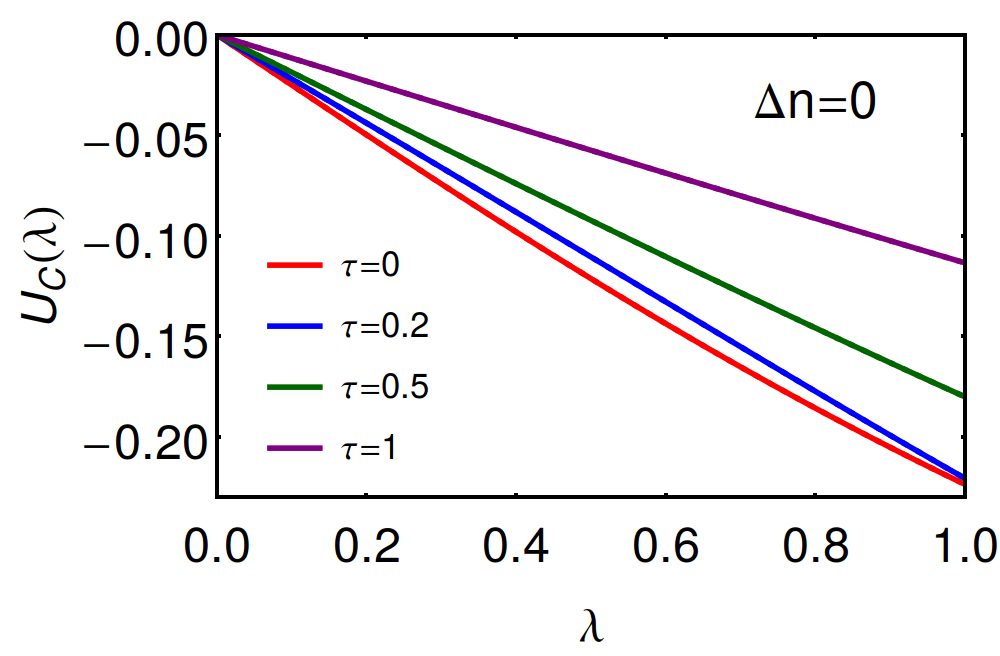}
\caption{Adiabatic connection integrand for the symmetric dimer at several
different temperatures.}
\label{Ac}
\end{figure}

\begin{figure}[htb]
\includegraphics[width=\columnwidth]{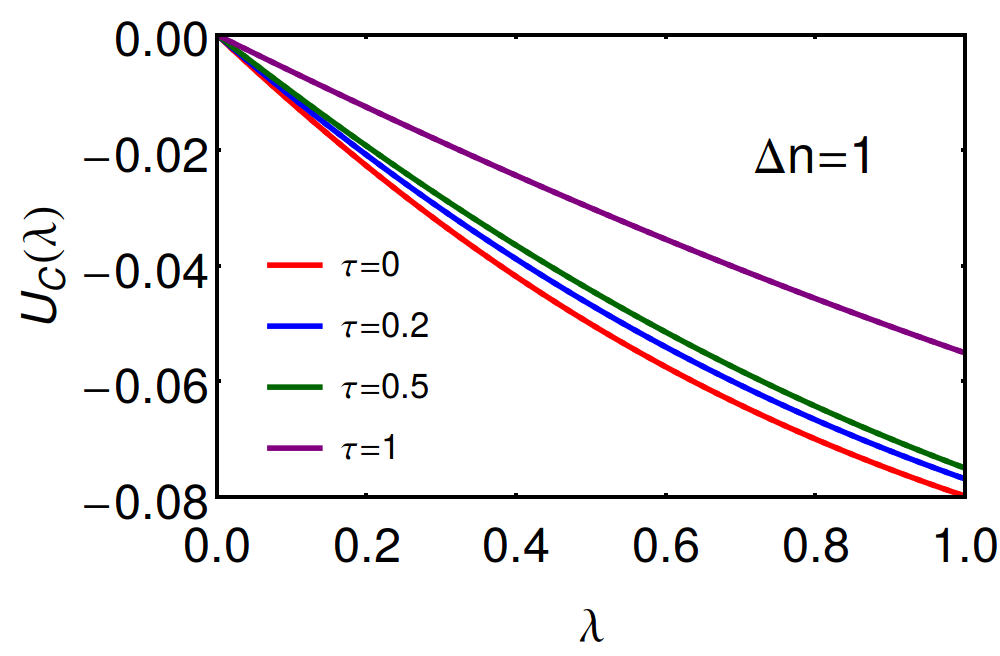}
\caption{Same as Fig. \ref{Ac}, except $\dn=1$.}
\label{Acdv2}
\end{figure}

Next, we consider the adiabatic connection formula\cite{LP75,GL76} that has
proven useful in studying and improving density functional approximations.
The ground-state version was calculated for the Hubbard dimer in Fig. 21 of
Ref. \cite{CFSB15}.  An alternative version, called the thermal connection
formula, was derived in Ref. \cite{PB15}, but that flavor relies on relating
the coupling-constant to coordinate scaling.  Such a procedure applies to
continuum models, but not lattices.  So we use the traditional version here,
applied to finite temperature\cite{PPFS11}:
\ben
A\c\t[\n]=\int_0^1 \frac{d\lambda}{\lambda} U\c\tl[\n]
\label{ACF}
\een
where $\lambda$ is a coupling constant inserted in front of $\hat V\ee$ in
the Hamiltonian, but (unlike regular many-body theory) the density is held fixed
during the variation.  Here $U\c\tl$ is the potential correlation energy at
coupling constant $\lambda$, which, for our model, is obtained by replacing $U$
with $\lambda U$.
In Fig. \ref{Ac}, for the symmetric case, turning on temperature
clearly reduces both the magnitude of the correlation and the degree of
static correlation, as judged by the initial slope of the curves.  
Fig. \ref{Acdv2} shows this result remains true beyond the
symmetric case.

\begin{figure}[htb]
\includegraphics[width=\columnwidth]{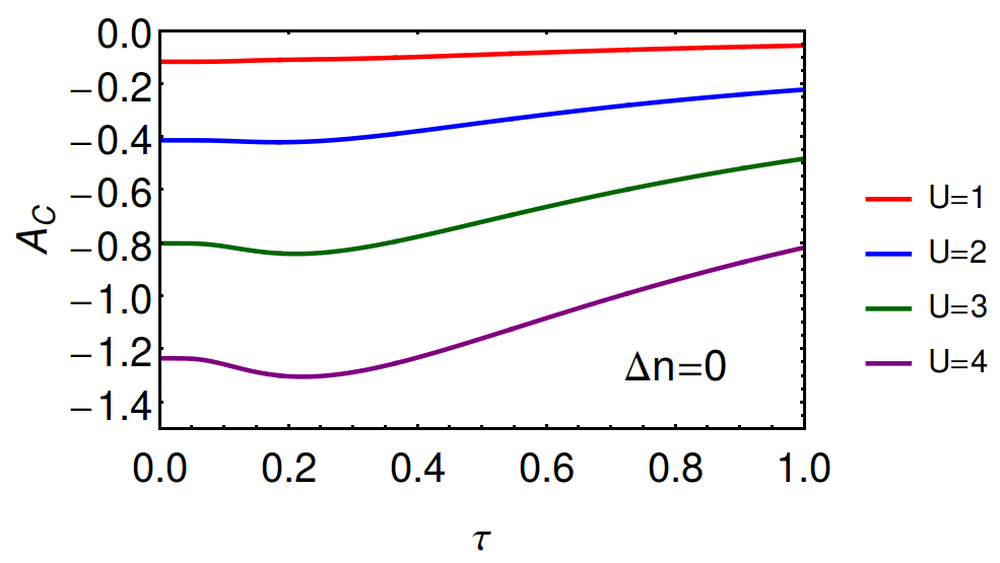}
\caption{Correlation free energy for the symmetric case with increasing values of $U$ ranging from weak to strong correlation. 
}
\label{AU5}
\end{figure}

In Fig. \ref{AU5}, we repeat the $A\c\t$ curves of Fig. \ref{corr} but now for fixed $\dn=0$ and 
increasing $U$. We start with the $U=1$ from earlier and increase into the strongly correlated
regime. The curves show a minimum at about $\tau=0.25$, particularly in $U=3$ and 4. Thus the
derivative with respect to temperature can be negative, and this does not happen even if we look closely at 
 $U=1$. Thus the correlation free energy is not generally monotonically decreasing in magnitude and the correlation energy is not bounded by the $\tau=0$ value.

\sec{Zero-temperature approximation}
\label{ZTA}

In this section, we explore the effects of making
the zero-temperature approximation (ZTA), in which thermal contributions
are ignored (Eq. (\ref{eqZTA})).  We use the (essentially) exact parametrization of
the ground-state XC energy of the Hubbard dimer of Eq. (108) of Ref. \cite{CFSB15}.
This substitution is made in the calculation of the total
free energy and in the MKS equations via the calculation of the XC potential, Eq. (\ref{xcpot}).
We return to Fig. \ref{AU1}, where we also plot the free energy in the ZTA by replacing $A\c^\tau(\dn)$ with $E\c(\dn)$, evaluated on the self-consistent $\dn$. 
We see that the error of ZTA is extremely small for $\tau \lesssim 0.5$.  Moreover, trends are very well
reproduced by the ZTA values, and fractional errors shrink for large $\tau$.  
This suggests that free energies in such calculations may be reliable depending, of course, 
on the precision needed in a given calculation.  
The errors grow most rapidly with $\tau$ when the dimer is asymmetric.
Thermal effects push the electrons apart, making the density more symmetric, 
in direct competition with $\dv$. For larger $\dn$, there is
a larger error in ignoring thermal effects.
Note that since we have only two electrons, our model is a worst
case scenario.  In many simulations, there are more valence electrons
per site, and (exchange-)correlation components are a much smaller fraction
of the total energy.  In a realistic DFT calculation, the error made
by approximating the ground-state functional would likely be much larger than
the error due to the lack of temperature-dependence\cite{SD14}.

\begin{figure}[htb]
\includegraphics[width=\columnwidth]{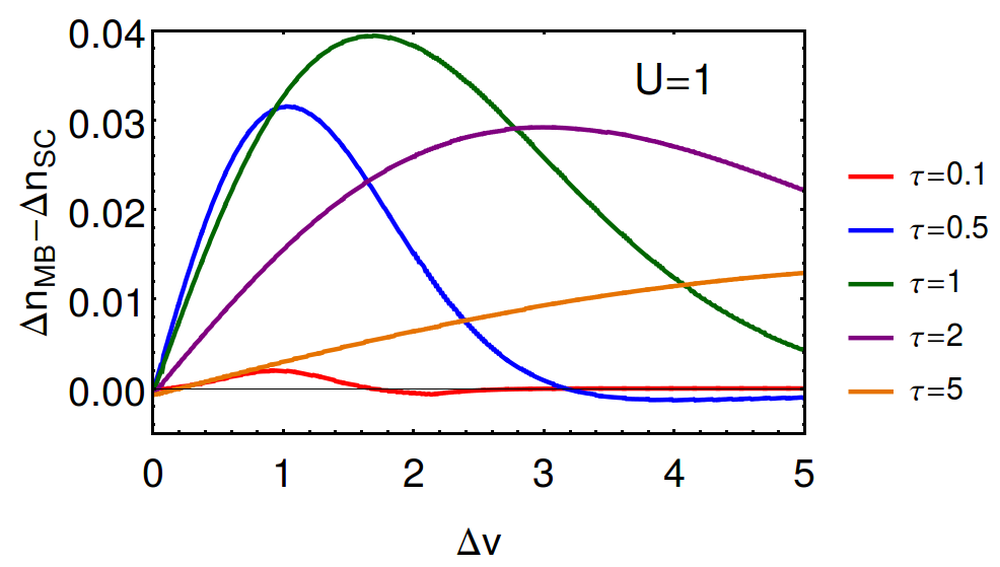}
\caption{Error in ZTA densities of Fig. \ref{dens}, density from self-consistent MKS subtracted from exact density.}
\label{dendiff}
\end{figure}
However, this is only part of the story.  Real thermal DFT calculations are performed
self-consistently within ZTA.  Then both the density and MKS orbitals are
often used to calculate response properties (usually on the MKS orbitals)\cite{S99,SAP97,RRCN02,CRLN02,DKC02,RCRN03,D05,RLDC09,PSTD12}. 
In Fig. \ref{dens}, we compare the self-consistent density
obtained using Eq. (\ref{MKSeqn}) through Eq. (\ref{ndef}).
In Fig. \ref{dendiff}, we plot the differences.
We see that the maximum errors in the density are small. 
At first they grow with small temperature but quickly start to
lessen as temperature increases which will be further explained below. 
As $\dv$ gets large the error goes to zero since the asymmetry dominates over
thermal effects.

In terms of Fig. \ref{corr}, the ZTA consists of approximating each of the curves by the 
corresponding black one.  Because all correlation components tend to vanish
with increasing temperature, while the total free energy grows in magnitude,
the small error made in the ZTA becomes less relevant with increasing temperature.
Specifically, we can analyze the symmetric case where correlation effects
are at their strongest. At $\tau=0$ correlation is about 20\% of the total energy but
when the system is at $\tau=1$ correlation is roughly 2.5\% of the total free energy. More importantly
this is due to the total energy magnitude going up by a factor of 5 and the correlation only decreasing by
a factor of 2.
This explains the small errors in the ZTA free energies of Fig. \ref{AU1} and 
the behavior of the self-consistent ZTA densities of Fig. \ref{dendiff}.
Note that the temperatures need not be so high as to make the density
uniform (i.e., symmetric).  Fig. \ref{dens} shows that, even for the temperature
at which density differences can be largest ($\tau=1$), the density difference
can remain substantial as the temperature increases, if the inhomogeneity ($\dv$) is
large enough.

\sec{Conclusions}

In summary, we have solved the simplest possible non-trivial system at finite
temperature exactly, both for the many-body case and within MKS density
functional theory.  We have produced the first 
exact plots of MKS quantities and the ZTA approximation
for a finite system (albeit one with a limited Hilbert space).
When the system is weakly correlated system at low to moderate temperatures, 
the neglect of thermal contributions to the
exchange-correlation functional has relatively little effect on the 
calculated free energies and even less on the self-consistent densities.  
Present limitations of ground-state approximations, such as their
inability to treat strongly correlated systems, are likely the greatest
source of error in these calculations. Future work will explore other quantities
of interest within thermal DFT and will analyze the ZTA more deeply.

\acknowledgments

The authors acknowledge support from the U.S. Department of Energy (DOE), Office of Science, Basic Energy Sciences (BES) under award \# DE-FG02-08ER46496.
J.C.S. acknowledges support through the NSF Graduate Research fellowship program
under award \# DGE-1321846.
A.P.J. acknowledges support through the DOE Computational Science Graduate Fellowship, grant number DE-FG02-97ER25308, and from the University of California Presidential Postdoctoral Fellowship Program.  Part of this work was performed under the auspices of the U.S. Department of Energy by Lawrence Livermore National Laboratory under Contract DE-AC52-07NA27344.

\appendix*

\sec{Energies and Densities for all States}
\label{app}

Here we list all the total energies, energy components, and density components for all 
particle numbers so that all the relevant ensemble averages of Eq. (\ref{ensavg}) can be reconstructed.
We begin with the energies
\bea
E_i(4-N) &=&  (2-N) U + E_i(N) ~~~~~~ N=0\rm{\textendash}4, \nonumber\\
E_0(0) &=& 0,  \nonumber \\
E_{0,1}(1) &=& \mp\sqrt{(2\,t)^2+\dv^2}/2, \nonumber \\
E_i(2) &=& \frac{2\,U}{3} - \frac{2\,r}{3} \cos(\theta + \frac{2\pi}{3} (i+1)) ~~~~~~ i=0,1,2, \nonumber\\
E_i(2) &=& 0, ~~~~~~ i=3,4,5, \nonumber 
%E_{0,1}(3) &=& U + E_{0,1}(1), \nonumber \\
%E_0(4) &=& 2\,U, \nonumber 
\eea
where
\bea
r &=& \sqrt{3((2\,t)^2+\dv^2)+U^2},\nonumber\\ 
\theta &=& \frac{1}{3}\arccos\left[\frac{9U(\dv^2 -2\,t^2) - U^3}{(3((2\,t)^2+\dv^2)+U^2)^{3/2}}\right]\nonumber.
\eea
$E_1(2)$ and $E_2(2)$ are both positive and should be ordered 4 and 5 instead. However the three triplets,
 i.e. the three zero-energy states, give only zero values in the later expectation values, so for notational convenience we order the non-zero 2-particle states 0, 1, and 2 instead of 0, 4, and 5.
These energies were used to construct $Z_{gc}$ in Eq. (\ref{zgc}) of the main text.

Next are the expectation values needed to construct the three different ensemble averages of interest, $T$, $V\ee$, and $\dn$ ($V\ext$ is unnecessary since it is trivially $\dv\dn/2$):
\bea
T_i(4-N) &=& T_i(n) ~~~~~~ N=0\rm{\textendash}4\nonumber\\
%T_0(0) &=& 0, \nonumber\\
T_{0,1}(1) &=&  \mp \frac{t}{\sqrt{(2\,t)^2+\dv^2}},  \nonumber\\
T_i(2) &=& (\beta_i^+ + \beta_i^-)^2/E_i(2) ~~~~~~ i=0,1,2,\nonumber
%V_{\rm{ee},0}(0) &=& V_{\rm{ee},0}(1) = V_{\rm{ee},1}(1) =0, \nonumber \\
\eea
\bea
V_{\rm{ee},i}(4-N) &=& (2-N)U + V_{\rm{ee},i}(N) ~~~~~~ N=0\rm{\textendash}4, \nonumber\\
V_{\rm{ee},0,1}(1) &=& 0,   \nonumber \\
V_{\rm{ee},i}(2) &=& U ((\beta_i^+)^2 + (\beta_i^-)^2) ~~~~~~ i=0,1,2, \nonumber
\eea
\bea
%\dn_0(0) &=& 0, \nonumber\\
\dn_i(4-N) &=& \dn_i(N) ~~~~~~ N=0\rm{\textendash}4, \nonumber\\
\dn_{0,1}(1) &=&  \mp \frac{2\dv}{\sqrt{(2\,t)^2+\dv^2}},\nonumber\\
\dn_i(2) &=& 2((\beta_i^-)^2 - (\beta_i^+)^2) ~~~~~~ i=0,1,2, \nonumber
\eea
and all the 0-particle terms are 0.
The $\beta^\pm$'s are from the $N=2$ wavefunction:
\ben
|\Psi_i(N)\rangle = \alpha_i(N) (|12\rangle + |21\rangle) + \beta_i^+(N)|11\rangle + \beta_i^-(N) |22\rangle \nonumber
\een 
with
\bea
\alpha_i &=& \frac{2\,t\,(E_i(2) - U)}{c_i E_i(2)}, ~~ \beta_i^\pm = \frac{U-E_i(2) \pm \dv}{c_i}, \nonumber\\
c_i &=& \sqrt{2(\dv^2 + (E_i(2) - U^2)^2(1+(2\,t/E_i(2))^{2})}. \nonumber
\eea
The ket $|ij\rangle$ signifies an electron at site $i$ and site $j$.
These expectation values were used with Eq. (\ref{ensavg}) to construct the densities and energy
components shown in the figures.

\bibliography{Master,Masterold,hubbard,thermal}
\label{page:end}
\end{document}